\documentclass[12pt]{iopart}
\usepackage{iopams}  
\usepackage{graphicx}  
\begin{document}
\title{Status of Chemical Freeze-Out}
\author{
{\sc J. Cleymans$^a$, H. Oeschler$^b$, K. Redlich$^c$, S. Wheaton$^a$}
}
\address{
$^a$ UCT-CERN Research Centre and Department  of  Physics, \\
Rondebosch 7701, Cape Town, South Africa\\
$^b$ Darmstadt University of Technology, D-64289 Darmstadt, Germany\\
$^c$ Institute of Theoretical Physics, University of Wroc{\l}aw,\\
 Pl. Maksa Borna 9, 50-204  Wroc{\l}aw, Poland\\
}
\ead{cleymans@qgp.phy.uct.ac.za}

\begin{abstract} 
The status of 
the energy dependence of the chemical freeze-out temperature 
and chemical potential obtained in heavy ion collisions 
is presented.
Recent proposals for chemical  freeze-out conditions are 
compared.
\end{abstract}

\maketitle

\section{Introduction}
\begin{figure}
\centering
\includegraphics[width=\linewidth]{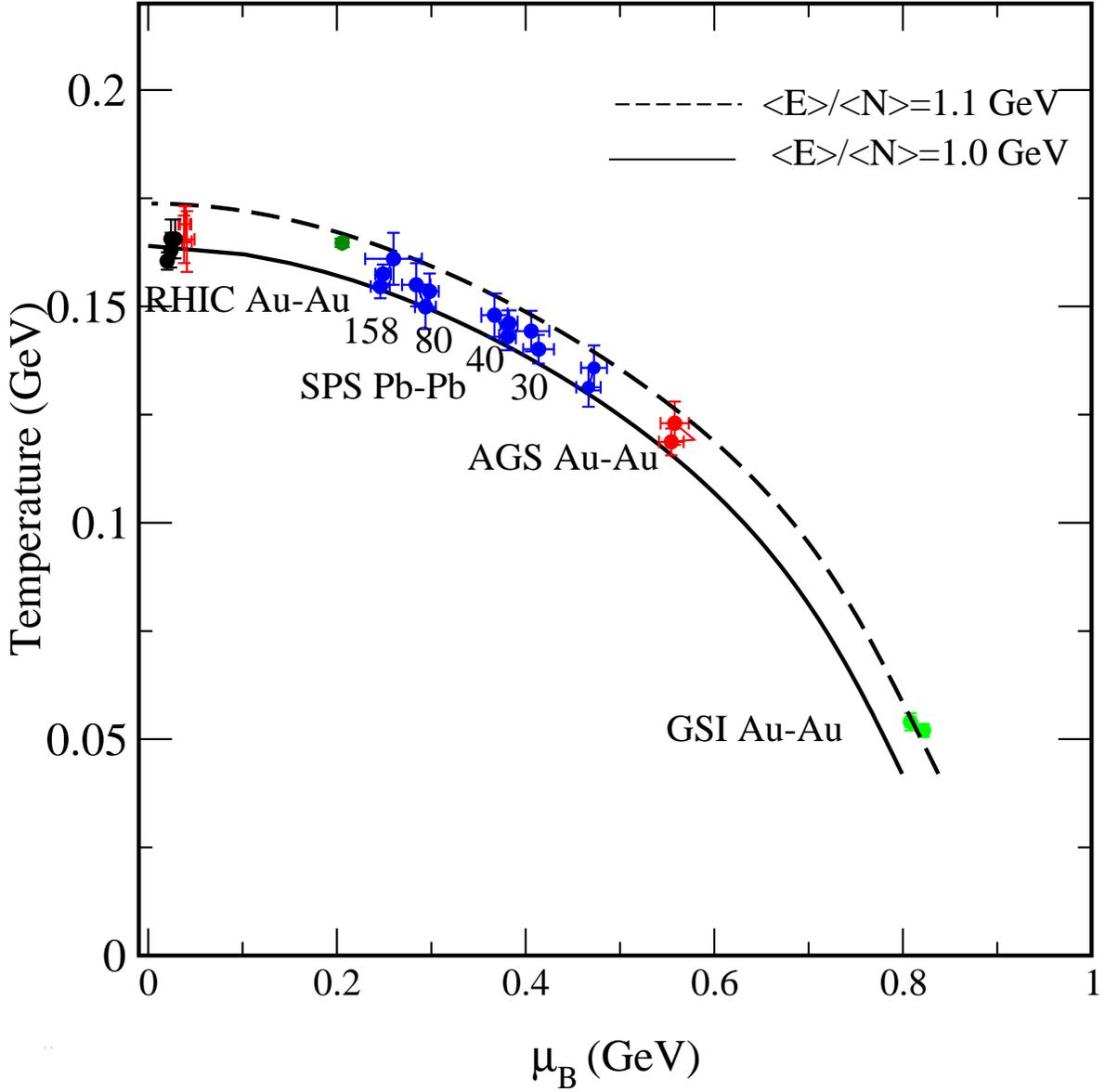}
\caption{\label{eovern06}Values of $T$ and $\mu_B$ deduced from particle multiplicities in heavy ion 
collisions for a wide range of beam energies.}
\end{figure}

Over the past  decade a   striking regularity has been established
in heavy ion collisions: from 
SIS  to  RHIC, particle yields are consistent with the assumption of
chemical equilibrium~\cite{review}. 
Furthermore, the chemical freeze-out temperature, $T$,  and the baryon
chemical potential $\mu_B$ follow a strikingly regular  pattern
as the beam energy increases.
This has led to several 
proposals  describing the freeze-out curve in
$T-\mu_B$ plane. 
The conditions of  fixed  energy
per particle~\cite{redlich1,redlich2}, baryon+anti-baryon density~\cite{pbm}, 
normalized entropy
density~\cite{horn,tawfik}  as well as percolation model~\cite{satz} all 
lead to 
reasonable descriptions of the freeze-out curve in the $T-\mu_B$ plane.
The
results have been compared with the most 
recent~\cite{manninen,andronic,xu} chemical freeze-out
parameters obtained in the thermal-statistical  analysis of
particle yields in~\cite{prc06} where 
the  sensitivity and dependence of the results on
parameters is analyzed and discussed. It has been shown in~\cite{prc06}
that, within present accuracies,
all chemical freeze-out criteria give a fairly good description of
the particle yields, however, the low energy heavy-ion data favor
the constant energy per hadron as a condition for
chemical  freeze-out. This condition also shows the
weakest sensitivity on model assumptions and  parameters.
%%%%%%%%%%%%%%%%%%%%%%%%%%%%%%%%%%%%%%%%%%%%%%
%
This  criterion  was first identified~\cite{redlich1,redlich2} by 
comparing the thermal parameters at
SIS energy with those obtained at SPS.  It was shown that the
average energy per particle  at SIS energy reaches  approximately
the same value of 1 GeV as calculated at the critical temperature
expected for deconfinement at $\mu_B=0$. In addition, known
results for chemical freeze-out parameters at the AGS also
reproduced the same value of energy per particle.  Thus, it was
suggested that the condition of a fixed energy per hadron is the
chemical freeze-out criterion in heavy-ion collisions. 
A comparison with the extracted results on $T$ and $\mu_B$  is shown in 
Fig.~[\ref{eovern06}]. The best estimate gives a value 
$\left<E\right>/\left<N\right>\approx$ 1.08 GeV.

\begin{figure}
\centering
\includegraphics*[width=\linewidth]{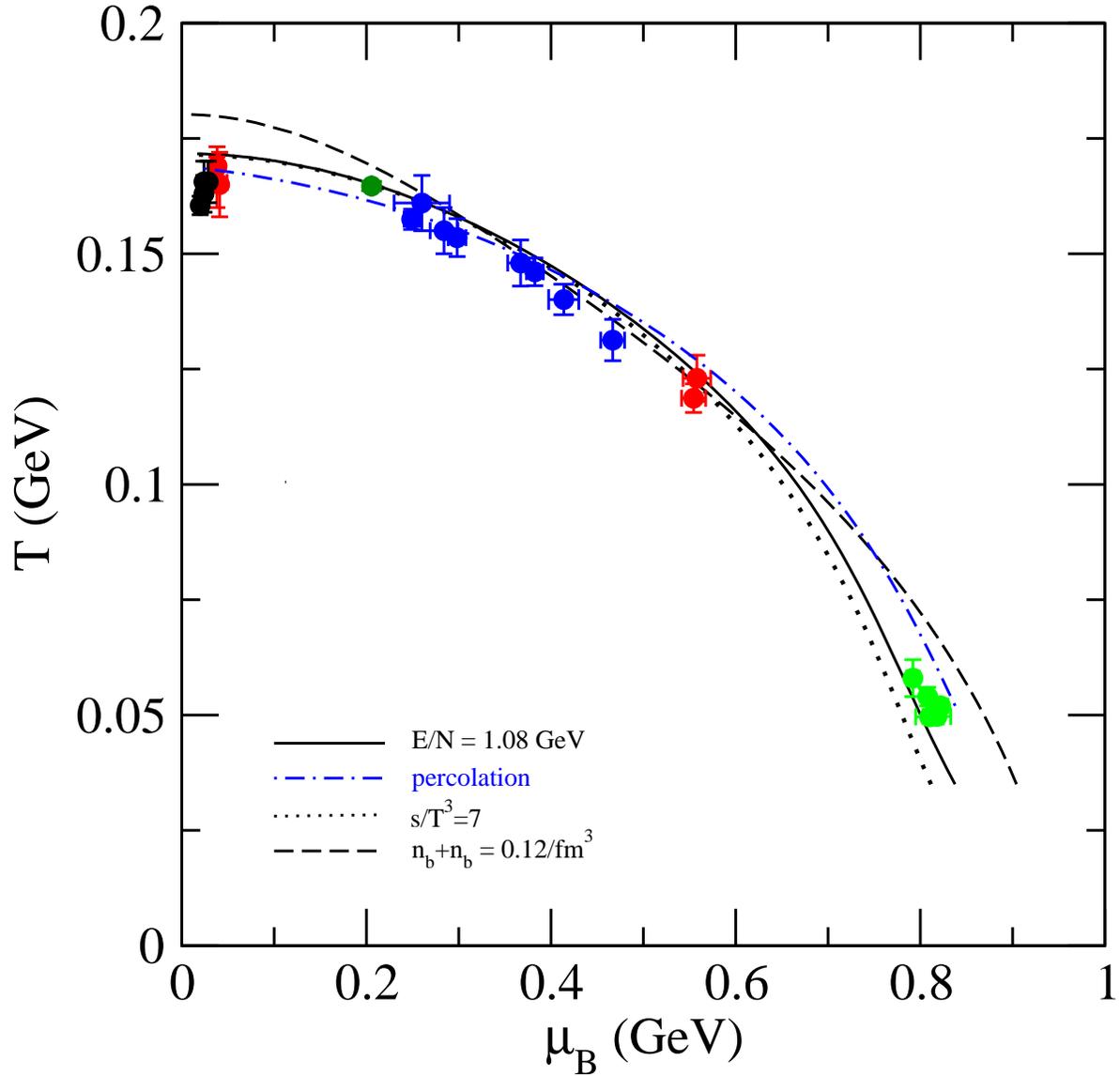}
\caption{Comparison of various freeze-out criteria with the values of $T$ and $\mu_B$
obtained from particle multiplicities in heavy ion collisions.}
\label{comparison}
\end{figure}

In addition to the fixed  $\left<E\right>/\left<N\right>$ criterion,  alternative
proposals have been made  to describe chemical freeze-out in
heavy-ion collisions at all
energies:
\begin{itemize}
\item a fixed value for the sum of
baryon and anti-baryon densities, $n_B+n_{\bar{B}}$, of
approximately 0.12/fm$^3$~\cite{pbm};
\item a self-consistent
equation for the densities based on geometric estimates using
percolation theory~\cite{satz};
\begin{equation}
n(T,\mu) = \frac{1.24}{V_h}
\left[1-\frac{n_B(T,\mu)}{n(T,\mu)}\right]+\frac{0.34}{V_h}
\left[\frac{n_B(T,\mu)}{n(T,\mu)}\right].
\end{equation}
\item a fixed value of the entropy
density, $s/T^3$, of approximately
7~\cite{horn,tawfik}.
\end{itemize}
A comparison of these proposals  is given in Fig.[\ref{comparison}]
which shows that all proposals give a reasonable description 
in the region between AGS and RHIC energies. Deviations appear at the highest
RHIC energy and at beam energies between AGS and SIS.
It would therefore be very interesting to have good data in this energy region.

\begin{figure}
\includegraphics*[width=\linewidth]{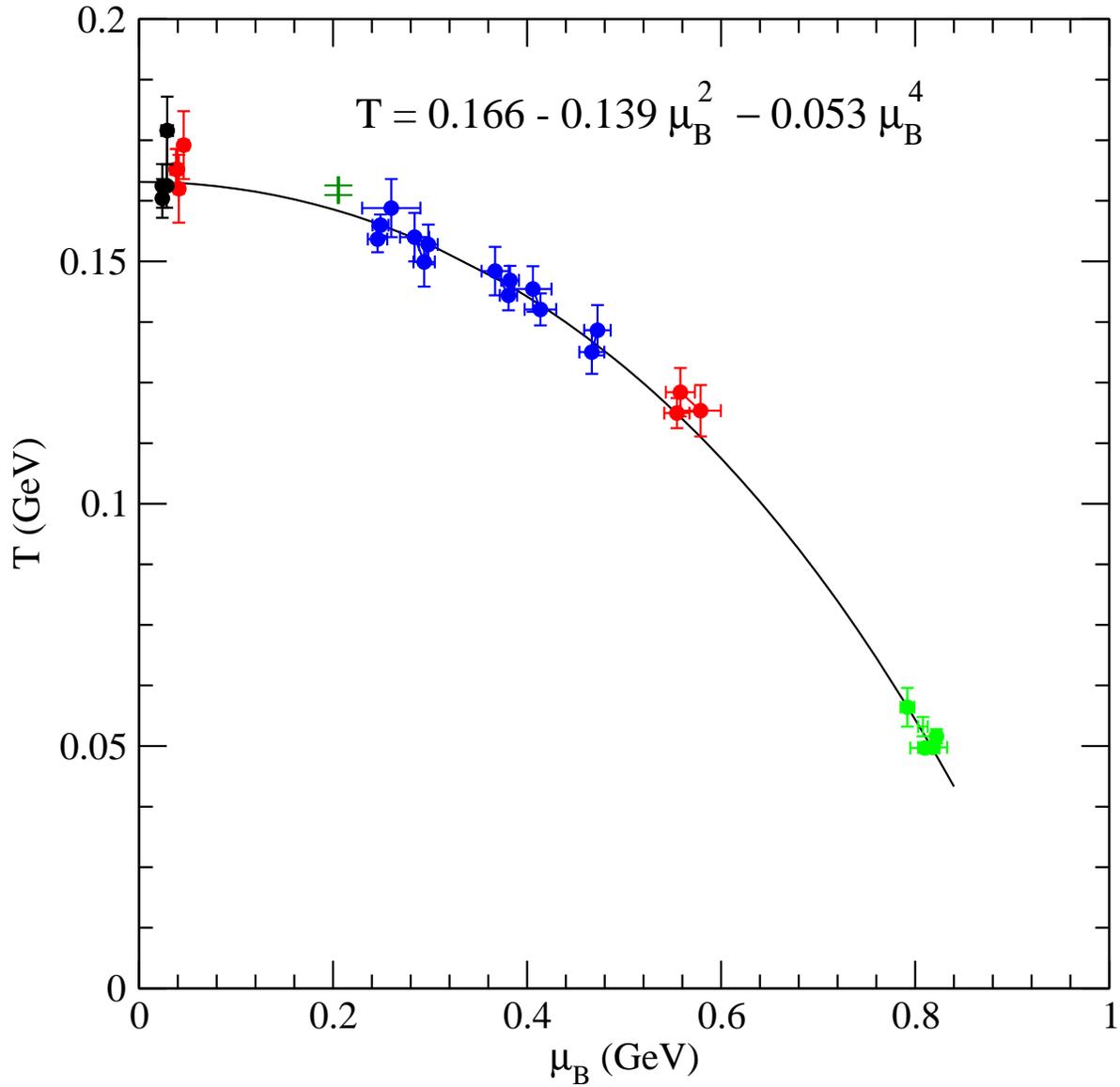}
\caption{A parametrization of the freeze-out curve deduced from 
particle multiplicities in heavy ion collisions.}
\label{numeric}
\end{figure}

Independently of any particular criterium or model for the freeze-out condition,
a numerical parametrization, shown in Fig.~[\ref{numeric}], is given by.
\begin{equation}
T = 0.166 - 0.139 \mu_B^2 -0.053\mu_B^4 .
\end{equation}
%
%
%
%%%%%%%%%%%%%%%%%%%%%%%%%%%%%%%%%%%%%%%%%%%%%%%%%%%%%%%%%
%
%%%%%%%%%%%%%%%%%%%%%%%%%%%%%%%%%%%%%%%%%%%%%%%%%%%%%%%%%
%%%%%%%%%%%%%%%%%%%%%%%%%%%%%%%%%%%%%%%%%%%%%%%%%%%%%
%%%%%%%%%%%%%%%%%%%%%%%%%%%%%%%%%%%%%%%%%%%%%%%%%%%%%%%%%%%
%
%
%%%%%%%%%%%%%%%%%%%%%%%%%%%%%%%%%%%%%%%%%%%%%%%%%%%
\section{Energy Dependence of $\mu_B$ and $T$.}
\begin{center}
\begin{figure}
\includegraphics*[width=\linewidth]{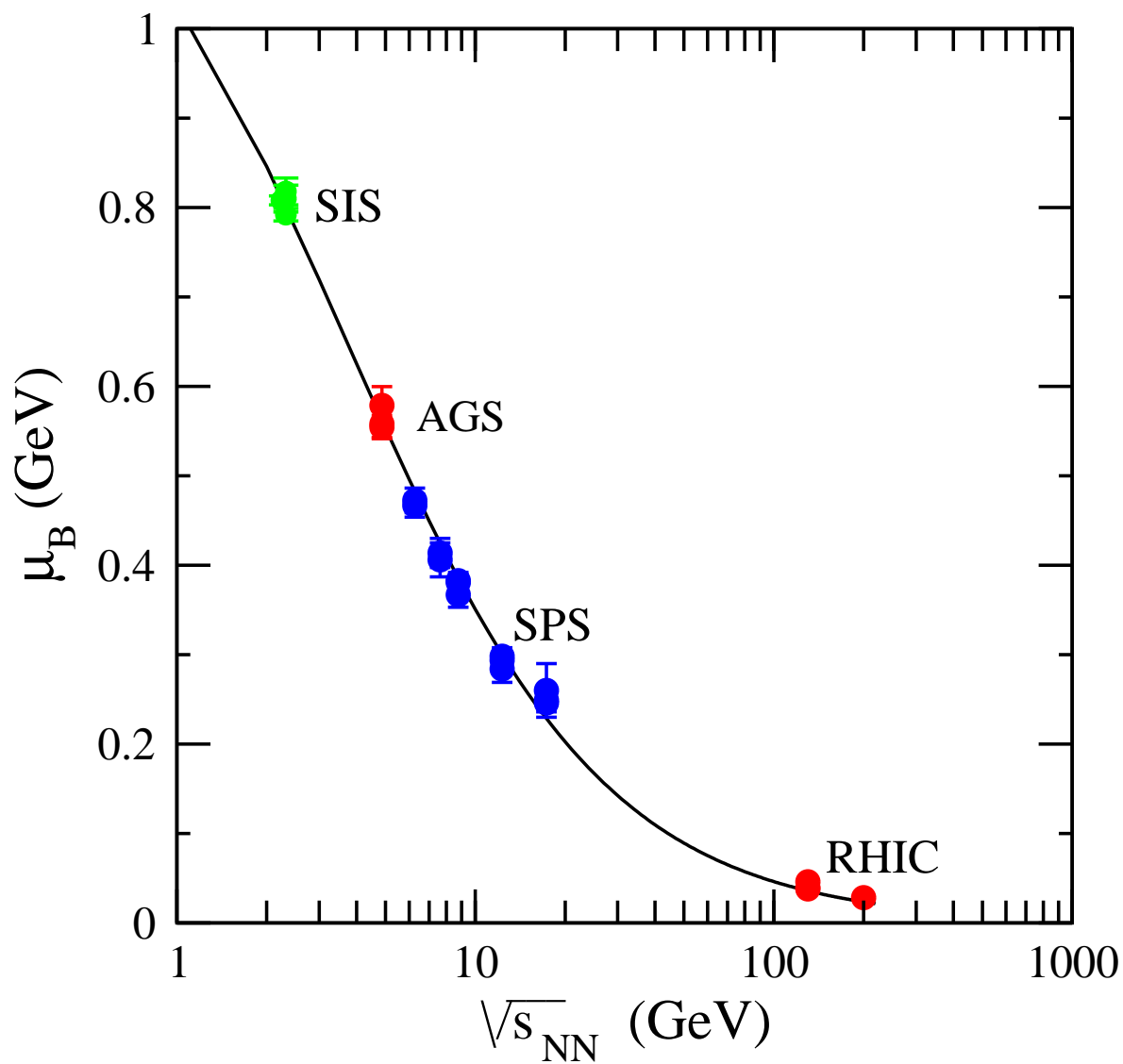}
\caption{Variation of the baryon chemical potential as a function of energy.}
\label{e_mub}
\end{figure}
\end{center}

\begin{center}
\begin{figure}
\includegraphics*[width=\linewidth]{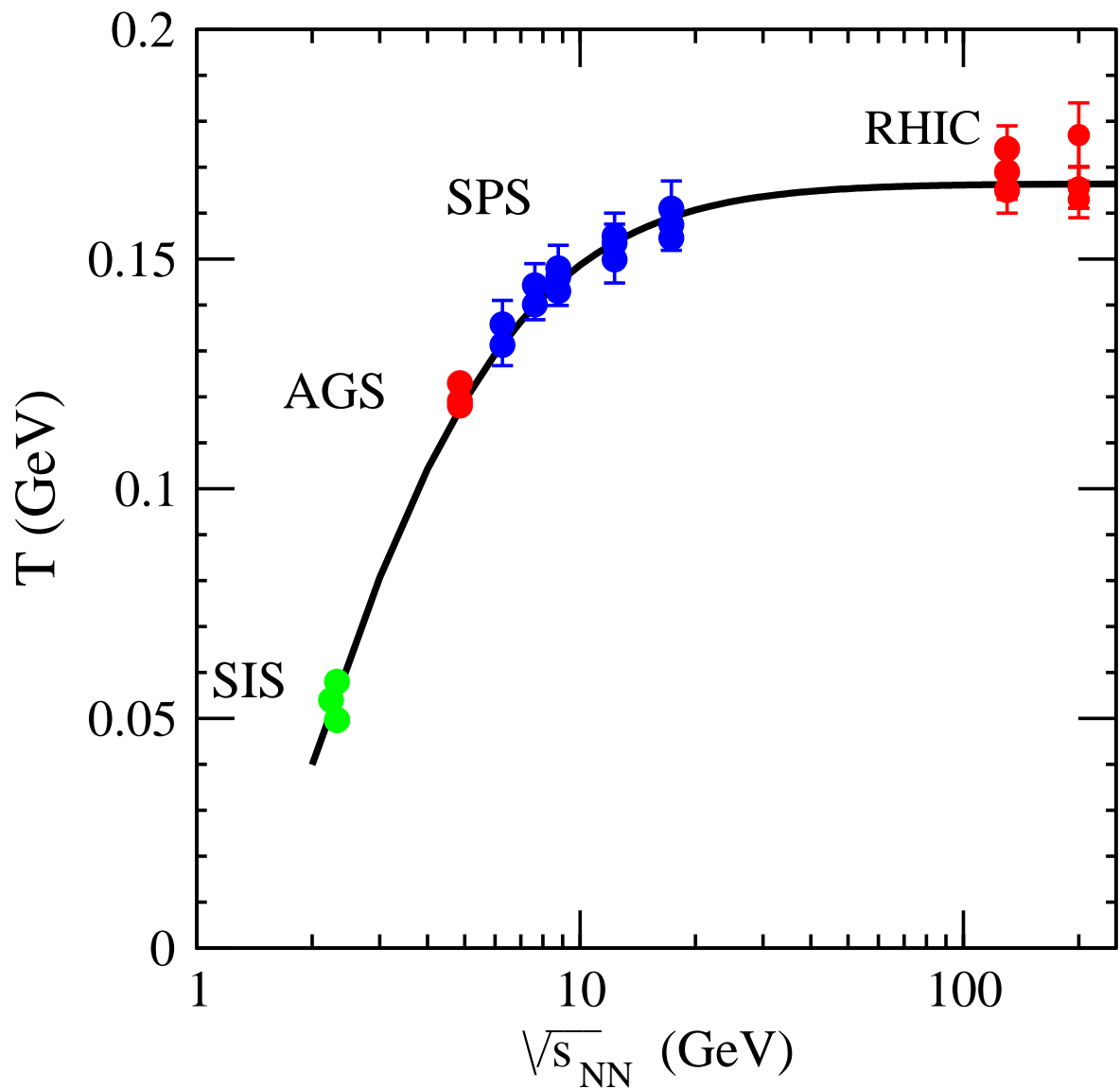}
\caption{Variation of the temperature as a function of energy.}
\label{e_t}
\end{figure}
\end{center}

The values obtained for $\mu_B$ as a function of beam energy 
are displayed in Fig.~[\ref{e_mub}]. As  this shows a smooth variation with
energy, it can be parametrized as
\begin{equation}
\mu_B(\sqrt{s}) = \frac{1.308~\mathrm{GeV}}{1 + 0.273~{\mathrm{GeV}}^{-1}\sqrt{s}}.
\end{equation}
This leads to the expectation that  $\mu_B\approx 1$~MeV at LHC energies.
%%%%%%%%%%%%%%%%%%%%%%%%%%%%%%%%%%%%%%%%%%%%%%%%%%%%%%%%%
Similarly, the freeze-out temperature  is shown in Fig.~[\ref{e_t}]. 
A straightforward extrapolation leads to a value at 
LHC energies  $T\approx 166$~MeV~\cite{kraus}.
%%%%%%%%%%%%%%%%%%%%%%%%%%%%%%%%%%%%%%%%%%%%%%%%%%%%%%%%%%%
%
%
\section{Conclusions}
There is by now a long history of measurements of particle abundances in 
heavy ion collisions covering a wide range of beam energies. The 
case  for chemical equilibrium has become stronger 
over the years with every new analysis confirming and reinforcing  conclusions
reached previously. To distinguish between the various proposals which have been made in 
the literature, the lower energy range 
at the AGS  acquires a special significance as it will
make it possible to discriminate between them. 
\section*{References}
%===================================
 
%-----------------------------------------------------------
%
\end{document}